\documentclass[11pt]{book}
\usepackage{Wiley-AuthoringTemplate}
\usepackage{chapterbib}
\usepackage[sectionbib,authoryear]{natbib}

\usepackage{amsmath}
\usepackage{graphicx,times,bm,url}
\usepackage{afterpage,lscape}
\graphicspath{{./fig/}{./png/}}
\usepackage{color}
\let\tablenum\undefined
\usepackage{siunitx}

\DeclareSIUnit\parsec{pc}
\DeclareSIUnit\lightyear{ly}
\DeclareSIUnit\year{yr}
\DeclareSIUnit\gauss{G}

\begin{document}

\graphicspath{{./fig/}{./png/}}
\let\tablenum\undefined


\newcommand{\AAA}{\mbox{\boldmath $A$} {}}
\newcommand{\uu}{\mbox{\boldmath $u$} {}}
\newcommand{\BB}{\bm{B}}
\newcommand{\JJ}{\mbox{\boldmath $J$} {}}
\newcommand{\DD}{{\rm D} {}}
\newcommand{\nab}{\mbox{\boldmath $\nabla$} {}}
\def\cs{c_{\rm s}}
\newcommand{\FF}{\mbox{\boldmath $F$} {}}
\newcommand{\SSS}{\mbox{\boldmath $S$} {}}
\newcommand{\dd}{{\rm d} {}}
\newcommand{\ee}{\mbox{\boldmath $e$} {}}
\newcommand{\Fig}[1]{Figure~\ref{#1}}
\newcommand{\gggg}{\mbox{\boldmath $g$} {}}
\newcommand{\SSSS}{\mbox{\boldmath ${\sf S}$} {}}
\def\cp{c_{p}}
\def\cv{c_{v}}


\chapter{Stability of plasmas through magnetic helicity}

\author*[1]{Simon Candelaresi}
\author[2,3]{Fabio Del Sordo}

\address[1]{\orgdiv{School of Mathematics and Statistics},
\orgname{University of Glasgow},
\postcode{G12 8QQ},
\city{Glasgow},
\country{United Kingdom}}

\address[2]{\orgname{Institute of Space Sciences (ICE-CSIC), Campus UAB},
\street{Carrer de Can Magrans s/n},
\postcode{08193},
\city{Barcelona},
\country{Spain}}

\address[3]{\orgdiv{Osservatorio Astrofisico di Catania},
\orgname{INAF},
\street{via Santa Sofia, 78},
\postcode{95125},
\city{Catania},
\country{Italy}}

\address*{Corresponding Author: Simon Candelaresi; \email{simon.candelaresi@gmail.com}}

\maketitle

\begin{abstract}{Abstract}
Magnetic helicity, and more broadly magnetic field line topology,
impose constraints on the plasma dynamics.
Helically interlocked magnetic rings are in a non-trivial topological state.
It is harder to bring them into
a topologically trivial state than two rings that are not linked.
This particular restriction has the consequence that helical plasmas
exhibit increased stability in laboratory devices, in the Sun
and in the intergalactic medium.
Here we discuss how a magnetic field is stabilizing the plasma
and preventing it from disruption by the presence of magnetic helicity. 
We present observational results, numerical experiments and analytical results that illustrate
how helical magnetic fields strongly contribute to the long-term stability of some plasmas.
We discuss several cases, such as that of solar corona, toroidal fusion devices, the galactic and
extragalactic medium, with a special emphasis on extragalactic bubbles.
\end{abstract}

\section{Introduction}

In the context of magnetohydrodynamics (MHD), the physics of magnetized fluids, 
helical magnetic fields are known to play a special role.
They are important both in the context of magnetic field amplification
\citep[e.g.][]{BS05, 2016MNRAS.461..240B}, and for the characterization of the saturation stage
of magnetic field evolution in MHD instabilities  \citep[e.g.][]{2011PhRvE..84b5403C, BBDSM12}.
In this work we  discuss the importance of helicity conservation law \citep[e.g.][]{1987ApJ...319..207B}
for the evolution and the stability of plasmas.
Large-scale magnetic helicity is known to stabilize plasmas in different contexts.
One remarkable example is that of rising flux tubes below the solar photosphere.
Their rise is aided by the presence of an internal twist within the raising and
the erupting plasma patches \citep{Canfield1996b}.
With the absence of a twisted magnetic field, the rising flux tubes
are more prone to disruption as they rise through the convective zone.
With a topologically trivial and non-helical internal field
CMEs would disintegrate before reaching the solar photosphere.

There are two formalisms that tell us how magnetic helicity acts as
a stabilizer of the system.
One is the near perfect helicity conservation in most astrophysical systems
with their high magnetic Reynolds numbers.
The other is the realizability condition \citep{ArnoldHopf1986}
which poses a lower bound of the magnetic energy in presence of helicity.

However, helicity acts not only as stabilizer.
As magnetic fields are wound up and braided continuously,
they may be subject of an MHD instability, the kink instability \citep[e.g.][]{2013SoPh..287..185B},
generated by a strong twist of the magnetic field lines.
The field in the solar atmosphere can get rather twisted (e.g. \cite{Mei2018}).
However, based on theoretical evidence, \citep{Aschwanden-2019-874-131-AAS} argued that
for active regions the induced twist may be too low to lead to a kink instability.

Here we will give an account how magnetic helicity plays and important role in
the plasma's stability, particularly in the context of astrophysical magnetic fields.
Those range from small scales found in fusion devices up to kiloparsec
sized plasmas in the intergalactic medium.

In this paper we will start by illustrating in Sec.\ \ref{sec: helicity_conservation}
how the magnetic helicity conservation works.
Then we will deal in Sec.\ \ref{sec: equilibrium_state} with the final equilibrium state that should be
reached by plasmas, and with the constraints
that play a role in reaching this equilibrium (Sec.\ \ref{sec: field_relaxation_constraints}).
In Sec.\ \ref{sec: plasma_stability} we will see some examples of plasmas 
which reach stability as consequence of the existence of large-scale magnetic helicity,
whilst we will concentrate on examples of
plasmas on galactic and intergalactic scales in Sec.\ \ref{Galactic}.
In Sec.\ \ref{Bubbles} we will then move to a specific example, 
that of intergalactic cavities, for which a mechanism of stabilization through
large-scale magnetic helicity has been recently proposed. 
Finally, we will draw some conclusions in Sec.\ \ref{Conclusions}.

\section{Helicity Conservation}
\label{sec: helicity_conservation}

The conservation of magnetic helicity in high magnetic Reynolds number
regimes can be easily derived from the magnetohyrodynamics equations
for a resistive, viscous compressible gas
(e.g. \cite{Biskamp-2003-Magnetohydrodynamic_Turbulence}):
\begin{equation}
\frac{\partial \AAA}{\partial t} = \uu \times \BB - \eta\JJ,
\label{eq: dAdt}
\end{equation}
\begin{equation}
\frac{\DD \uu}{\DD t} = -\cs^{2} \nab \ln{\rho} +
\JJ\times\BB/\rho + \FF_{\rm visc},
\label{eq: dUdt}
\end{equation}
\begin{equation}
\frac{\DD \ln{\rho}}{\DD t} = -\nab \cdot \uu,
\label{eq: drhodt}
\end{equation}
with the magnetic vector potential $\AAA$, velocity $\uu$,
magnetic field $\BB = \nab\times\AAA$, magnetic resistivity $\eta$,
isothermal speed of sound $\cs$, density $\rho$,
current density $\JJ = \nab\times\BB$,
viscous forces $\FF_{\rm visc}$
and Lagrangian time derivative
$\DD/\DD t = \partial/\partial t + \uu\cdot\nab$.
Here the viscous forces are given as $\FF_{\rm visc} = \rho^{-1}\nab\cdot2\nu\rho\SSS$,
with the kinematic viscosity $\nu$, and traceless rate of strain tensor
$\SSS_{ij} = \frac{1}{2}(u_{i,j}+u_{j,i}) - \frac{1}{3}\delta_{ij}\nab\cdot\uu$.
Being an isothermal gas we have $p = \cs^{2}\rho$ for the pressure.
Alternatively to the induction equation in vector potential form
we could write it using directly the magnetic field $\BB$ as
\begin{equation}
\frac{\partial \BB}{\partial t} = \nab\times(\uu \times \BB) + \eta\nabla^{2}\BB.
\label{eq: dBdt}
\end{equation}

Here we use the induction equation \eqref{eq: dAdt} expressed using the magnetic vector
potential $\AAA$ instead of the magnetic field $\BB$.
This has a few advantages.
First, we can directly compute the magnetic helicity density
\begin{equation}
h_{\rm m} = \AAA\cdot\BB
\end{equation}
in all space for all times without the need of finding the inverse curl.
Second, the solenoidal condition of $\nab\cdot\BB = 0$
is trivially fulfilled.
This is particularly useful in numerical calculations as it is ensured
not to incur in non-zero-divergence $\BB$, which could potentially lead to spurious
results \citep[see e.g.][]{BrandScann2020ApJ...889...55B}.
Of course, we are still left with the gauge choice here,
since the vector potential $\AAA' = \AAA + \nab\phi$, with the differentiable
scalar field $\phi$, leads to the same vector field $\BB = \nab\times\AAA'$.
The implicit gauge in our induction equation is the the Weyl gauge
with vanishing electrostatic potential.

The helicity density depends on space and time.
For helicity conservation we are interested in the total helicity,
i.e.\ the volume integrated magnetic helicity density
\begin{equation}
H_{\rm m} = \int_V \AAA\cdot\BB\ \dd V.
\end{equation}
We will assume that the volume that we are interested in will remain static.
In a later discussion we will contrast the magnetic helicity
with the magnetic energy
\begin{equation}
E_{\rm m} = \frac{1}{2}\int_V \BB\cdot\BB\ \dd V.
\end{equation}

From the MHD equations we can now derive the rate of change in time
of the magnetic helicity as
\begin{equation}
\partial_t H_{\rm m} = \int_V \partial_t \AAA\cdot\BB + \AAA\cdot\partial_t\BB.
\end{equation}
Using the induction equation for $\AAA$ \eqref{eq: dAdt} and $\BB$ \eqref{eq: dBdt}
and $\nab\times\BB = \JJ$ we find
\begin{eqnarray}
\partial_t H_{\rm m} & = & \int_V (\uu \times \BB + \eta\nabla^{2}\AAA)\cdot\BB \ \dd V 
 + \int_V \AAA\cdot(\nab\times(\uu \times \BB) + \eta\nabla^{2}\BB) \ \dd V \nonumber \\
 & = & -2\eta\int_{V}\JJ\cdot\BB\ \dd V +
 \int_{\partial V}\AAA\times(\BB\times\uu - \eta\JJ)\cdot\dd\SSS.
\end{eqnarray}
The first term is the resistive helicity dissipation/generation term.
For turbulent dissipative systems, the electric current density and
magnetic field are on average aligned.
This makes this term {\em in practice} a dissipation term.
The last two terms are surface terms.
They have the potential of helicity injection or helicity loss in a finite domain.
For instance, this is the case for domains located in the solar atmosphere or bordering the solar photosphere.
On the photosphere this helicity injection can be substantial
(e.g.\ \cite{Linan-Pariat-2018-865-52-ApJ, Gupta-Thalmann-2021}).
In these cases, the fields are finite and there is a possible helicity flux through the surface, even for small
diffusivities, as the diffusivity does not enter the surface terms.
The same concept applies on large scales, such as galaxies, who may experience helicity loss through
the halo and then to extragalactic medium.
This helicity flux may be relevant for galactic dynamo \citep{2020A&A...641A.165N}
and may play a key role for galactic magnetic fields to reach the observed
values \citep{2013MNRAS.429.1686D, 2021arXiv210812037R}.

For systems with periodic boundaries, closed boundaries, or where the boundaries are so far that
our quantities can be considered zero, the surface terms vanish.
In that case we can focus on the resistive dissipation term.
This dissipation is small for most astrophysical settings, as $\eta$ is so small that
structures of the size in question diffuse on much longer time scales than the typical dynamical time scales
(e.g.\ \cite{Berger1984}).
So, we can assume that magnetic helicity is conserved in such systems.

In the limit of vanishing magnetic diffusion ($\eta = 0$) the induction
equation \eqref{eq: dAdt} simplifies to
\begin{equation}
\frac{\partial \AAA}{\partial t} = \uu \times \BB,
\end{equation}
or, expressed in terms of the magnetic field $\BB$, we can write
\begin{equation}
\frac{\partial \BB}{\partial t} = \nab\times(\uu \times \BB).
\end{equation}
This last equation describes a magnetic field that
is simply Lie-transported under the velocity field $\uu$ \citep{mimetic14}.
We than say that the field is ``frozen in'' into the fluid
\citep{Alfven-1942-150-405-Nature, BatchelorFrozeIn1950RSPSA,
PriestReconnection2000}.
In simple terms, a gas/fluid compression perpendicular to $\BB$
increases the local magnetic field strength by the compression factor.
We can also show that every open advected surface $S$ preserves
its magnetic flux
\begin{equation}
\Phi = \int_S \BB\cdot\dd S.
\end{equation}
This can be expressed in terms of differential two-forms (e.g.\ \cite{Arnold-TopologicalMethodsinHydrodynamics-2013, mimetic14})
$\beta = B_x\dd y \wedge \dd z + B_y\dd z \wedge \dd x + B_z\dd x \wedge \dd y$
\begin{equation}
\frac{\dd}{\dd t}\beta = \mathcal{L}_{\uu}(\beta),
\end{equation}
where $\mathcal{L}_{\uu}(\beta)$ is the Lie-transport of $\beta$ under
the velocity $\uu$.

As the magnetic field is frozen in,
magnetic field lines obtain a physical meaning, as they cannot be
broken up and reconnected, which is a manifestation of the restrictions
in the field's dynamics.

Together with the magnetic field, also the magnetic helicity density is being Lie-transported
in the ideal limit.
Being a density, this means that every advected volume $V$ preserves
its helicity.
Consequently, also the entire domain (being also a volume) conserves its magnetic helicity:
\begin{equation}
\frac{\dd}{\dd t}H_{\rm m} = 0.
\end{equation}
So, we have exact magnetic helicity conservation for $\eta = 0$.

The discussion is similar for the case of the limit, i.e.\ $\lim_{\eta \rightarrow 0}$.
We can perform the same calculations and arrive at the same result
\begin{equation}
\lim_{\eta \rightarrow 0} \frac{\dd}{\dd t}H_{\rm m} = 0.
\end{equation}

\section{Relaxation Equilibrium State}
\label{sec: equilibrium_state}

What consequences does the helicity conservation have for the dynamics of the system,
particularly for relaxing/decaying magnetic fields?
During any stage of its evolution -- this of course includes its end state --
it must go through states of the same magnetic helicity.
This already excludes a large class of fields, like those with a connectivity
with a different magnetic helicity.

So, similar to the energy conservation in a non-dissipative system like a pendulum,
we can already make predictions about what behavior is excluded and, to some extend,
make qualitative predictions about the geometry of the magnetic field lines
during the evolution.

For a field in a non-equilibrium state, we know that it will try to reach
a minimum energy state.
This is true for any relaxing magnetic fields in a closed or infinitely
large MHD system.
Without any constraints, this lowest energy would simply be zero.
\cite{Woltjer-1958-489-91-PNAS} showed that under magnetic helicity
conservation, the lowest magnetic energy state is the force-free state
with $\nab\times\BB = \alpha\BB$, where $\alpha$ is the force-free parameter.
This state cannot always be reached under the dynamical time evolution.
This depends on the geometry of the magnetic field lines and their topology.
While magnetic helicity would pose restrictions, other topological
quantities may also affect the obtainable equilibrium state.
For instance, if the minimum energy state can only be reached by magnetic field line
reconnection, it won't be accessible under ideal evolution.

Since in the ideal ($\eta = 0$) case the field evolves under a Lie-transport,
every advected sub-volume of the domain conserves its helicity.
In particular, every finite neighborhood of the magnetic field lines,
conserves it helicity content.
For closed field lines of finite length this can be easily imagined.
We then simply have infinitesimally thin magnetic closed flux tubes
(potentially knotted or braided) that conserve helicity.
This is true for all neighboring flux tubes that may contain
a different amount of helicity.
For ergodic field lines, that fill a finite space in three dimensions,
the interpretation becomes more interesting.
In that case we have a constant helicity not just on a sub-volume with
measure $0$, but with potentially a finite measure and finite volume
(e.g.\ \cite{Arnold-TopologicalMethodsinHydrodynamics-2013}).

Under ideal conditions with no breaking and coalescing of magnetic
field lines (no reconnection), \cite{Taylor-1974-PrlE}
suggested that the helicity is conserved for every sub-volume
bounded by the magnetic field line.
That means that not only the total, but also the sub-helicities
are conserved.
The minimum energy state (relaxed state) is then a non-linear force-free field
where the force-free parameter $\alpha$ depends on the field line:
\begin{equation}
\nab\times\BB = \alpha(a, b)\BB.
\end{equation}
For ergodic field lines we then have finite volumes with the same parameter
$\alpha$, as we are dealing with the same field line.
However, under real conditions with magnetic reconnection
\cite{Taylor-1974-PrlE} further argued that it is only the
total magnetic helicity that is conserved, as sub-volumes bounded
by magnetic field lines do not make any physical sense.
In that case, the minimum energy state is a linear force-free state
of the form
\begin{equation}
\nab\times\BB = \alpha\BB,
\end{equation}
where $\alpha$ is a constant in the domain.

\section{Field Relaxation Constraints}
\label{sec: field_relaxation_constraints}

From the above discussion by Woltjer and Taylor we cannot deduce if the given
magnetic field will ever reach such an equilibrium.
It can well be that topological constraints inhibit the plasma to even get close
to a linear or non-linear force-free state.

One such constraint is given by the realizability condition \citep{ArnoldHopf1986}.
It defines a lower limit for the magnetic energy in presence of magnetic helicity as
\begin{equation}\label{eq: realizability}
2 E(k) \ge k|H(k)|,
\end{equation}
where $k$ is the inverse wavelength.
This can be integrated over the parameter $k$, which results
into a global constraint.
In an experiment or numerical simulation this effect can be easily
observed by comparing the magnetic helicity with the magnetic energy.
After some initial free relaxation with a drop in energy,
the energy reaches a near constant value in low resistivity environments.
Independent of the magnetic resistivity, the ratio of the two
quantities will reach a near constant.

For a purely hydrodynamical case that follows the compressible viscous
Navier-Stokes equations, an analogous exact relation exists.
But instead of relating the kinetic energy to the kinetic helicity
the relation is between the kinetic helicity and the enstrophy,
which is the volume integral of the vorticity squared.
It can be easily derived that the enstrophy is bound from below
by the kinetic helicity \citep{vortex_braids}.
Although with no existing exact relation, the same authors
found expirementally a similar realizability condition
that limits the kinetic energy from below by the presence of
unsigned kinetic helicity.

A na\"{i}ve interpretation of the realizability condition can be drawn from a simple thought experiment.
Take two closed magnetic flux tubes, with no internal twist, that are linked with each other once.
This configuration is helical.
A minimum energy state cannot be reached, without reconnection.
However, we can conceive an experiment with three linked fluxtubes
and by defining the correct signs of the magnetic field in each tube,
we can make this configuration non-helical.
This puts us into a slightly troubled situation, as the na\"{i}ve interpretation
of the realizability breaks down.
From numerical experiments by \cite{DS2010PhRvE}, we know that
the actual linking plays only a minor role, while the helicity content poses
the real restrictions, as expressed in the realizability condition.

Taylor relaxation can not always be realized.
Due to additional constraints of existing topological quantities, the proposed
equilibrium state cannot always be reached without breaking the field's topology.
For instance, \cite{Yeates_Topology_2010} showed that for a field containing
finite field line helicity the equilibrium state was not the Taylor state.

\section{Plasma Stability}
\label{sec: plasma_stability}
\subsection{Solar Eruptions}

Analyzing soft X-ray images of solar active regions, \cite{Canfield1999}
showed that sigmoidal structure are more likely to result into solar eruptions
compared to non-sigmoidal structures.
Sigmoids are a result of magnetic flux tube twisting where
an increase in twist leads to kinking.
These flux tubes are generated below the photosphere where they can obtain
their internal twist from the solar rotation (e.g.\ \cite{Canfield1996b}),
although already emerged flux tubes can get twisted through
photospheric footpoints where the tubes are anchored
(e.g. \cite{Candelaresi-2016-Corona}).
As they rise through the convection zone some of them disrupt and do
not pinch through the photosphere as an intact fluxtube.
Those that are helical, however, survive long enough and may
result into a coronal mass ejections at later times \citep{Gibson2002}.

While the twist leads to stability below the photosphere,
it can also lead to a kink instability, if it is sufficiently high
(e.g.\ \cite{Finkelstein-Weil-1978-17-201-IntJTheoPhys,
Craig-Sneyd-1990-357-653-ApJ,
RustKumar1996ApJ,
Ebrahimi-Karami-2016-162-1002-MNRAS,
Vemareddy-Gopalswamy-2017-850-38-ApJ}).
Such high fluxtube twists can be generated through photospheric motions
\citep{Canfield1999}.

At the same time, an induced twist into a more complex photospheric
magnetic loop structure has the potential to move
the fluxtubes around, such that reconnection may occur.
Even on a scale relatively small, such as that of sunspots,
helicity can be induced into fluxtubes via sunspot rotation
(e.g.\ \cite{Wang-Liu-2016-SolPhys}).
These reconnection events lead to a high electric current density
and with that a high particle acceleration.
Such events may not be coronal mass ejection, but rather bright flares
that can be observed in white light or in extreme ultra violet.

\subsection{Toroidal Fusion Plasma Fields}

A magnetically confined toroidal fusion device
is aimed at producing controlled fusion reactions in hot plasma
(e.g.\ \cite{wesson1997tokamaks}).
In order to confine the plasma in the shape of a torus, strong magnetic fields are used
(e.g. \cite{Taylor-1993-5-12-PhysFluB, Jardin-Ferraro-2015-115-21-PRL}).
A helical magnetic field in fusion plasma can easily maintain its energy:
this is evident from the realizability condition \eqref{eq: realizability}.
Such a helical field can be generated by two currents.
One is poloidal, i.e.\ a ring current around the torus' main axis.
The other component is a current along the main axis,
which can artificially injected.

From \cite{Taylor-1974-PrlE} and \cite{Taylor1986} we know that without
any energy input, the magnetic field relaxes into a force-free state
with $\nabla\times\BB = \alpha\BB$.
This is the minimum energy state under helicity conservation.
Therefore we can deduce the minimum energy input that is required to keep the plasma in a stable configuration
and hence allows to maintain it in a stationary state.

\subsection{Numerical Experimental Evidence}

In order to understand the mechanism behind the stabilizing effect
of the magnetic helicity in plasmas it is useful to conduct
a series of experiments.
Real world experiments do currently not have the power to measure
the mechanisms in sufficient detail.
But we need to know about the magnetic field relaxation mechanism, the
nature of the reconnection events and the energy conversion sites
(magnetic to kinetic/thermal).

A way out of this is by modelling the plasma using appropriate equations.
For a non-relativistic plasma, as we assume the solar, galactic and fusion plasmas
to be, we can make use of the MHD equations \eqref{eq: dAdt}-\eqref{eq: drhodt}.

From a purely geometrical interpretation \citep{MoffattKnottedness1969}
one can argue that linking (helicity)
is a stabilizer by preventing/restricting the flux tubes to reconnect arbitrarily.
An early numerical experiment was devised by \cite{DS2010PhRvE}.
They inserted three magnetic flux rings into a box (initial condition).
Two outer rings were linked to the inner ring, resulting into a linked configuration.
By changing the sign of the magnetic flux of one of the outer rings they were able to change
the system from one with net magnetic helicity to no magnetic helicity.
A control configuration with three un-linked flux tubes was also used.
With the non-helical field relaxing similarly to the trivial non-linked field
it was shown that it is the helicity, rather than the actual linkage that
restricts the field relaxation.

Similar experiments were carried out by \cite{C2011PhRvE}
where they studied the relaxation of helical and non-helical knots.
One of their results was that topologically non-trivial and non-helical
knots show a kind of intermediate stability between the topologically trivial
fields and the helical fields.
For non-helical braids \cite{Yeates_Topology_2010} observed similar
enhanced stability to the trivial case.
They argued that further topological invariants should be considered.
The consequences would be far ranging and have implications particularly for
the stability of toroidal fusion plasma fields.
\cite{2016A&A...587A.125P, 2016A&A...591A..16P} introduced a technique for generating
tubular magnetic fields with arbitrary axial geometry and internal topology
with the goal of investigating the behaviour of flux ropes whose field lines have more complex entangled configurations.
One of the main application of this technique is the study of flux ropes in solar corona. 
They concluded that magnetic field lines with sufficient complex entanglement can suppress large-scale morphological changes,
and magnetic energy can be reduced  through reconnection and expansion of the ropes.

\section{Galactic and Intergalactic Medium}
\label{Galactic}

\subsection{Observations}

Much of the discussion in the literature on helicity in plasmas focuses on the Sun
and to some lesser extent to fusion plasmas.
In order to explain the observed large-scale fields in galaxies, magnetic helicity
has been taken into account to model the magnetic field amplification through the dynamo effect.
Since the parameters are unsuitable for a direct numerical simulations, many authors
\citep{KleeorinMoss2000,
Sur-Shukurov-2007-377-874-MNRAS,
Shukorov-Sokolov-Subramanian-Brandenburg-2008,
ssHelLoss09,
Chamandy-Shukurov-2014-443-1867-MNRAS,
Chamandy-2016-462-4-MNRAS}
have used the mean-field approach in which the given fields are separated
into a fluctuating and a mean part, assuming there is scale separation.

While magnetic fields in galaxies are less studied than in stars and planets,
the field in the intergalactic medium is even less well explored.
From a number of X-ray observations we know that some galaxies in clusters, like
Cygnus A \citep{Carilli-Perley-1994-270-173-MNRAS} emit material
through jets that forms bubble like structures above the galactic plane.
The X-ray signature is a effect of the internal magnetic field
that can have strengths of $\SI{10}{\micro\gauss}-\SI{40}{\micro\gauss}$
\citep{Carilli-Taylor-2002-40-319-ARAA}.
Further observations by
\cite{Taylor-Fabian-2002-334-769-MNRAS,
Churazov-Bruggen-2001-554-261-ApJ,
Birzan2004ApJ,
McNamara-Nulsen-2007-45-117-ARAA}
and \cite{Montmerle-2011-51-299-EAS}
in different galaxy clusters, including the Virgo cluster with the galaxy M87,
suggest that these lobes, or bubbles, have a higher temperature
and lower density than the surrounding intergalactic medium
by a factor of ca.\ $3$ in both.
With a size of roughly $\SI{5}{\kilo\parsec}$, they are smaller, although
not significantly, than the galaxies from which they emanate.
Similar observations in other galaxies, the Milky Way too displays bubbles rising above and below its midplane.
They are known as Fermi bubbles, and they are known to emanate a multiwavelength radiation, from radio to $\gamma$-ray \citep{2015ApJ...799..112C}.
Possible mechanisms that lead to these bubbles are galactic fountain flows
\citep{Bregman1980ApJ},
the ejection of hot material through jets and the compression
of the intergalactic medium \citep{Churazov-Bruggen-2001-554-261-ApJ}.
Energy input from an active galactic nucleus can provide the necessary
power to heat the bubbles.

The plasma in the bubbles is magnetized and relatively hot.
Being hot and underdense they rise form the galactic plane much
like a hot air balloon, but without membrane.
This leads to a velocity shear at its boundary and should lead
to the Kelvin-Helmholtz instability.
Nevertheless, from measurements of their total life time (e.g.\ \cite{Birzan2004ApJ})
we know that they can survive for $\SI{10}{\mega\year}$-$\SI{100}{\mega\year}$).
This is much longer than what is estimated from the instability analysis.
So, apart from their origin, one of the puzzling aspects is
how they survive for such a long time.

\subsection{Kelvin-Helmholtz Instability}

In pure hydrodyanmics, a shearing flow with the velocity
$\uu = f(x)\ee_y$ can be unstable, depending on the shearing flow
profile $f(x)$ (typically its steepness) and the density profile of the gas
\citep{Chandrasekhar1961hhs}.
In its simplest form we can consider two phases.
One phase consists of a fluid or gas with density $\rho_1$ and constant (in space)
velocity $\uu_1$.
The other phase has density $\rho_2$ and parallel velocity $\uu_2 \ne \uu_1$.
Such a system, according to the Navier-Stokes equations, will not change in time.
Through the Kelvin-Helmholtz instability, however, a small perturbation will
be amplified exponentially, depending on the wave-length of the perturbation.

For a magnetized gas or plasma we can take the step into magnetohydrodynamics.
Here, the gas and the magnetic field are coupled through the Lorentz force.
Through magnetic tension building up through the fluid flow from the onsetting instability,
one could imagine a stabilizing effect from the Lorentz force.
\cite{Chandrasekhar1961hhs} studied the effect of a homogeneous external
magnetic field on the Kelvin-Helmholtz instability.
There are two simple cases.
One consists of a field that is perpendicular to the shearing flow ($\BB_0 \perp \uu$).
This case does not have any effect on the instability.
A parallel magnetic field, however, stabilizes the system.
The suppressed wavelengths depend on the strength of the field.
At a field strength of
\begin{equation}
B^2 \ge 2\pi(u_1 - u_2)^2(\rho_1\rho_2)/(\rho_1 + \rho_2)
\end{equation}
all modes are suppressed.

So, an external parallel magnetic field is a viable stabilizer for the buoyant
intergalactic bubbles.
Whether such a field exists or is strong enough is a matter to be
determined by the observers.
For comparison to our proposed stabilizing mechanism, we will discuss
this case in the next section.

\section{Helical Intergalactic Bubbles}
\label{Bubbles}
From helical galactic dynamo models (e.g.\ \cite{KleeorinMoss2000, Sur-Shukurov-2007-377-874-MNRAS}),
we can make the assumptions that the generated 
magnetic field is also helical.
If these bubbles originate from the galactic interior, from where they obtain their magnetic field,
we can also assume that this field is helical.
Being helical we can hypothesize that these bubbles have an intrinsic stability.

It is hard to make any precise predictions on the geometry and topology
of this internal field.
Therefore, we will discuss the effect of two different internal helical magnetic fields
and their effects on the stability of the intergalactic bubbles.
For more in-depth discussion and details see \cite{bubbles}.

\subsection{Numerical Experiment}

\begin{figure}
\begin{center}
\ifdefined\isgrey
  \includegraphics[width=0.3\columnwidth]{bubble_schematic_grey}
\else
  \includegraphics[width=0.3\columnwidth]{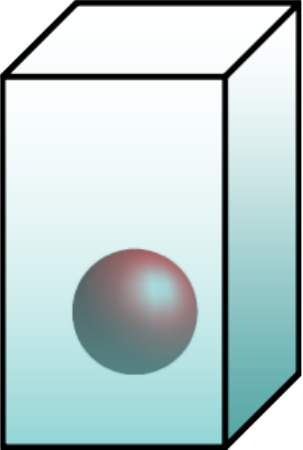}
\fi
\caption[]{
{
Schematic representation of the initial condition of the numerical experiment.
The hot underdense bubble with homogeneous temperature and density is
embedded in a stratified cold medium of higher density.
The galactic disk is at the lower boundary and exerts a constant gravitational force.
}
}
\label{fig: bubble_schematic}
\end{center}
\end{figure}

To test if a helical magnetic field can stabilize the intergalactic bubbles
such that they survive the observed \SI{10}-\SI{100}{\mega\year},
we performed a number of MHD simulations \citep{bubbles}.
The setup consists of a underdense hot bubble surrounded by a colder stratified medium
(see \Fig{fig: bubble_schematic}).
Buoyancy is generated by a homogeneous gravitational acceleration $g$.
Since the galactic disk is larger than the bubble, a constant gravitational pull
is a justified approximation and simplifies the numerical calculations.

In the direct numerical simulations we solve the full resistive, viscous and compressible
MHD equations with the energy equation, that includes the temperature.
The density and temperature contrast between bubble and medium is of a factor of $3$,
which is approximately what has been observed.
The numerical code used, the {\sc PencilCode}
\citep{Brandenburg-Dobler-2002-147-471-CompPhysComm, Brandenburg-2020-Zenodo}
(\url{https://github.com/pencil-code})
solves for the magnetic vector potential $\AAA$,
instead of the magnetic field $\BB = \nab\times\AAA$.
This makes sure that the solenoidal condition is fulfilled and facilitates the computation
of the magnetic helicity.
The equations we solve are
\begin{eqnarray}
\frac{\partial \AAA}{\partial t} & = & \uu\times\BB + \eta\nab^2\AAA,
\label{eq: induction} \\
\frac{\DD \uu}{\DD t} & = & -\cs^{2}\nab \left( \frac{\ln T}{\gamma} + \ln{\rho} \right)
 + \frac{\JJ\times\BB}{\rho} \nonumber \\
 & &  -\gggg + \FF_{\rm visc},
\label{eq: momentum} \\
\frac{\DD \ln{\rho}}{\DD t} & = & -\nab \cdot \uu,
\label{eq: continuity} \\
\frac{\partial \ln T}{\partial t} & = & -\uu\cdot\nab\ln T - (\gamma - 1)\nab\cdot\uu \nonumber \\
 & & + \frac{1}{\rho c_v T} \left( \nab\cdot(K\nab T) + \eta\JJ^2 \right. \nonumber \\
 & & \left. + 2\rho\nu\SSSS\otimes\SSSS + \zeta\rho(\nab\cdot\uu)^2 \right),
\label{eq: temperature}
\end{eqnarray}
with the magnetic vector potential $\AAA$,
magnetic field $\BB = \nab\times\AAA$,
fluid velocity $\uu$,
constant magnetic resistivity (diffusivity) $\eta$,
advective derivative $\DD / \DD t = \partial/\partial t + \uu\cdot\nab$,
sound speed $\cs = \gamma p/\rho$,
adiabatic index $\gamma = \cp/\cv$,
heat capacities $\cp$ and $\cv$ at constant pressure and volume,
temperature $T$,
density $\rho$,
electric current density $\JJ = \nab\times\BB$,
gravitational acceleration $\gggg$,
viscous force $\FF_{\rm visc}$,
heat conductivity $K$
and the bulk viscosity $\zeta$.
The viscous force is given as $\FF_{\rm visc} = \rho^{-1}\nab\cdot 2\nu\rho\SSSS$,
with the traceless rate of strain tensor $S_{ij} = \frac{1}{2}(u_{i,j} + u_{j,i}) - \frac{1}{3}\delta_{ij}\nab\cdot\uu$.
The equation of state used here is for the ideal monatomic gas and it appears
implicitly in our equations, as we eliminated pressure $p$.
Here the gas is monatomic with $\gamma = 5/3$.

\subsection{Helicity as Stabilizer}

Within our numerical bubble we insert a helical magnetic field.
To exclude effects from the geometry of the field we use to different setups.
One consists of the ABC type field of the form
\begin{equation}
\AAA = f(r) A_0 \left(
\begin{array}{c}
\cos((y-y_{\rm c})k) + \sin((z-z_{\rm c})k) \\
\cos((z-z_{\rm c})k) + \sin((x-x_{\rm c})k) \\
\cos((x-x_{\rm c})k) + \sin((y-y_{\rm c})k)
\end{array}
\right),
\label{eq: abc}
\end{equation}
where the wave number $k$ can be used to regulate the helicity relative to the
magnetic energy, which is regulated by the amplitude $A_0$.
Here the subscript c stands for cavity.
The function $f(r)$ only depends on the radius from the bubble's center and makes sure that
the field smoothly goes to zero at the boundary with no current layer forming.

\begin{figure}[t!]\begin{center}
\ifdefined\isgrey
  \includegraphics[width=0.7\columnwidth]{spheromak_grey}
\else
  \includegraphics[width=0.7\columnwidth]{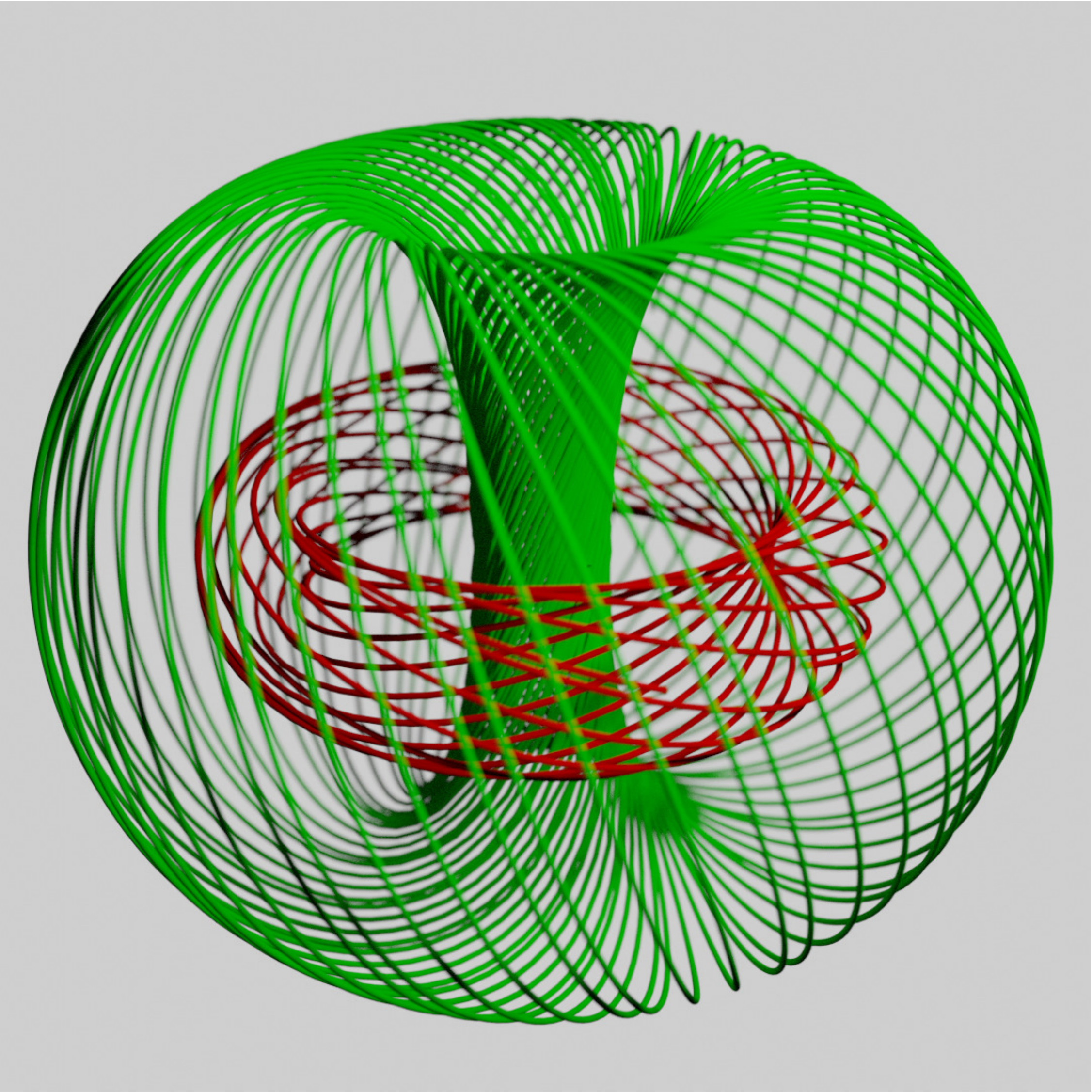}
\fi
\end{center}
\caption[]{
Helical spheromak configuration showing two of the magnetic field lines.
The original color image from \cite{bubbles} (under CC-BY 4.0 license) was monochromed.
}
\label{fig: spheromak}
\end{figure}

The second helical configuration is the spheromak configuration and
consists of twisted magnetic fields in a toroidal shape.
Those fields are embedded into each other, forming a highly helical field (see \Fig{fig: spheromak}).
One of the advantages of this configuration is that it smoothly approaches
zero field strength at the bubble's boundary.
Similar to the ABC case, we can adjust magnetic energy and magnetic helicity independently.
For a more detailed formulation see \cite{bubbles}.

For both cases we keep the magnetic energy constant while adjust the helicity
such that the high helicity case has four times the helicity of the low helicity case.
As control cases we also run simulations with no magnetic field and two
with only an external field.
This gives us seven different simulations.

The stability of the bubble is measured using a coherence measure $d_{\rm mean}$,
which is simply the mean distance of the points within the bubble.
Weather or not a point belongs to the bubble is determined by the tempearature.
All points above a threshold, by definition, belong to it.
So, a large value of $d_{\rm mean}$ tells us that the bubble has undergone
some significant disruption.

\begin{figure}[t!]\begin{center}
\ifdefined\isgrey
  \includegraphics[width=0.8\columnwidth]{coherence_TT_dmean_z_grey} \\
  \includegraphics[width=0.8\columnwidth]{coherence_TT_dmean_z_sph_grey}
\else
  \includegraphics[width=0.8\columnwidth]{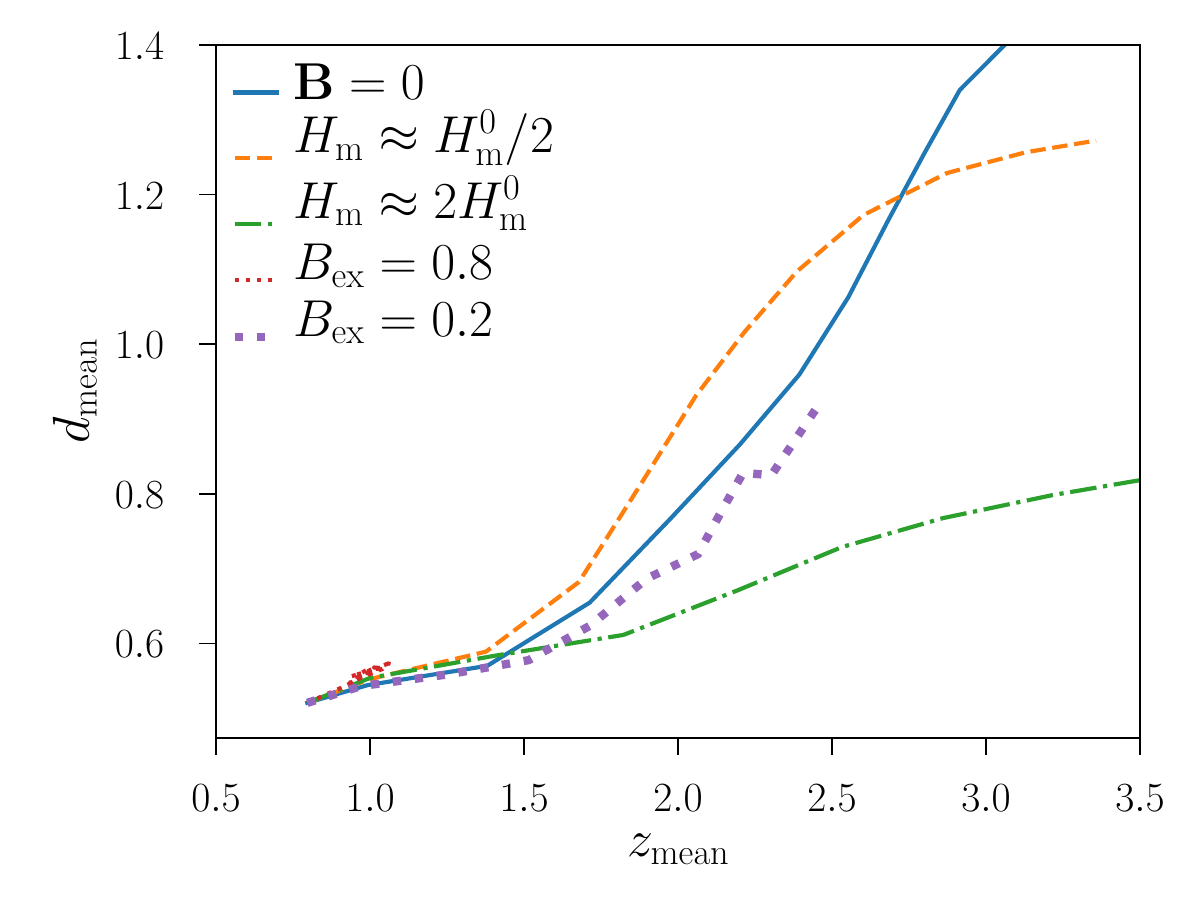} \\
  \includegraphics[width=0.8\columnwidth]{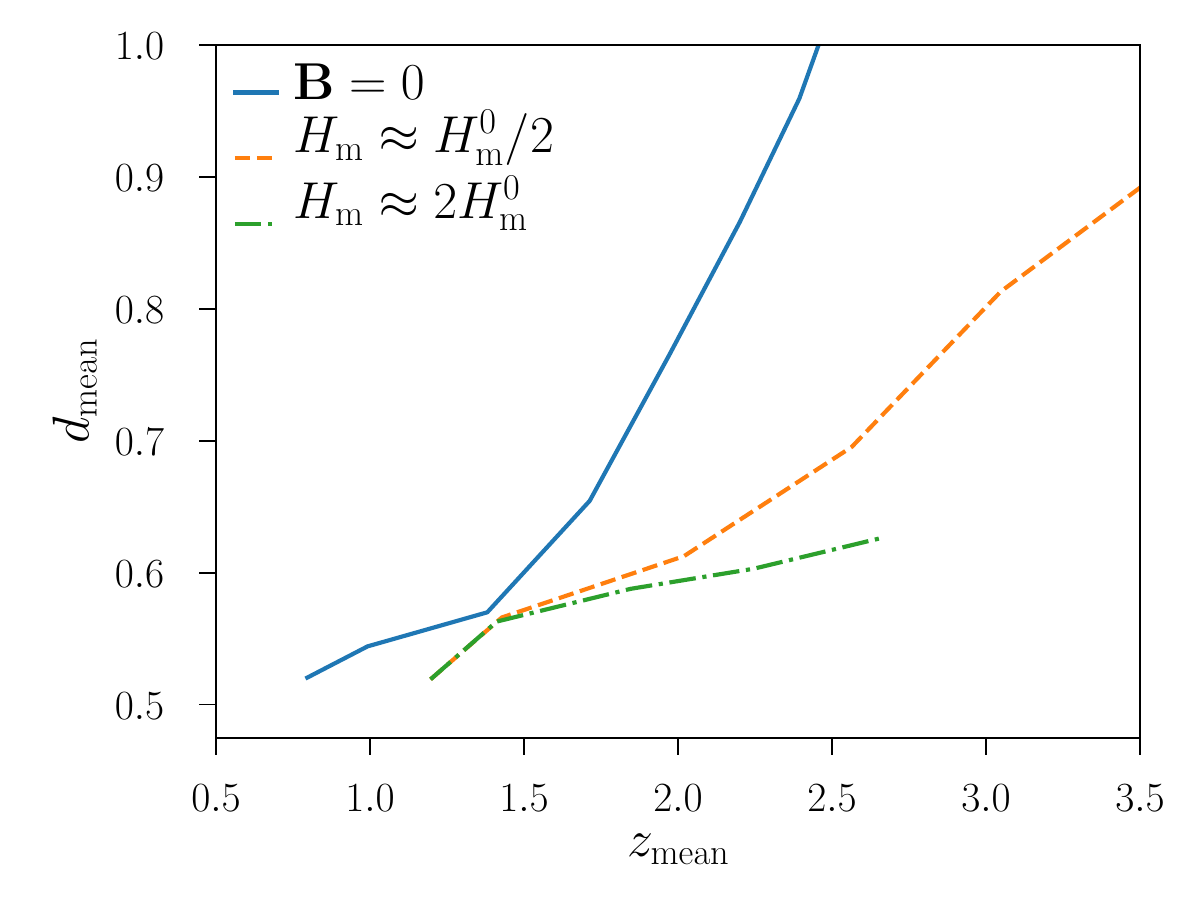}
\fi
\end{center}
\caption{
Coherence measure $d_{\rm mean}$ of the intergalactic bubbles in dependence of the mean height.
The upper panel shows the control case with no magnetic field ($\BB = 0$) with the
ABC field in two different helicity contents and same energy, and the case of an external
magnetic field with strength close to the predicted value for the suppression of the instability
on all wavelengths.
We notice that the red dotted line, corresponding to case $B_{ex}=0.8$,
is very short because the strong field prevents the bubble to raise.
The lower panel shows the case for the spheromak configuration.
The original color figures from \cite{bubbles} (under CC-BY 4.0 license) were monochromed.
}
\label{fig: coherence_TT_dmean}
\end{figure}

From our calculations (\Fig{fig: coherence_TT_dmean}) we clearly see that
a helical internal magnetic field stabilizes the field.
This is independent of the type of helical field used (ABC versus spheromak).
At a minimum level of helicity we reach stability.
If we want the same amount of stability from an external field,
however, we require a much larger total magnetic energy content.
We estimate that a helical field with maximum strength of the order of $\SI{e-5}\gauss$ can
stabilize the bubble over a time scale of about \SI{250}{\mega\year}.
For comparison, these simulations also show that in a purely hydrodynamical case the
bubbles can be stable for about \SI{80}{\mega\year}.

\section{Conclusions}
\label{Conclusions}

We have discussed the role of helical magnetic fields in the stabilization of plasmas.
First of all we have seen how magnetic helicity is a conserved quantity in ideal MHD and in
the high-conductivity limit, that is when magnetic diffusivity is very small.
This is the case for various astrophysical plasmas on very different length scales.

Helicity conservation is well illustrated by considering magnetic tubes which are interlocked.
Whenever their configuration is characterized by a finite magnetic helicity,
it results into a much more stable system than in non-helical cases.
On the other hand, the topological linkage of such tubes plays no role in stabilizing the plasma.
Therefore, magnetic helicity conservation imposes strong constraints on the relaxation
of magnetic fields whenever some helicity is present.

We have discussed helicity conservation in the stability of structures in
the solar corona, such as coronal mass ejections, as well as in fusion plasma fields.
On much larger scales, such as galactic and extragalactic environments, we see a similar behavior.
In particular, some extragalactic structures moving at high velocity through the
extragalactic medium, are observed to be surprisingly stable
towards disruption by the Kelvin-Helmoltz instability.
Magnetic helicity may, therefore, play an important role against sucha a distruption.

In the last section we have therefore examined the possibility that a helical
magnetic field may stabilize extragalactic bubbles even at low magnetic energy densities.
These bubbles are observed raising from the midplane of our galaxy and other galaxies
into the intergalactic medium.
We justify the presence of helicity in these bubbles by the fact the they are boing inflated by
an active galactic nucleus or from jets coming from the galactic center, which yield helical fields.

We tested two different types of helical magnetic fields: the ABC flow and a spheromak field.
This field initially sits inside the bubble raising through buoyancy in an otherwise stably stratified medium.
We quantified the disruption of the bubble and observed that this bubble is stabilized by the
helicity towards the Kelvin-Helmholtz instability in the non-linear regime
if the magnetic field is sufficiently helical.
Our estimate is that a helical field with maximum strength of the order of $\SI{e-5}\gauss$ can
stabilize the bubble over a time scale of at least \SI{250}{\mega\year},
which is more than three times longer than the stability of a non-magnetized bubble.
This remarkable example illustrates the role played by magnetic helicity in the
stabilization of structures in plasmas.
Moreover, it also suggests that helical magnetic fields are present in astrophysical
magnetized plasmas wherever stable configurations are observed, and throughout several scales,
from stars to extragalactic structures.

\section*{Acknowledgments}
Fabio Del Sordo acknowledges support from
a Marie Curie Action of the European Union (grant agreement 101030103),
the programme Unidad de Excelencia María de Maeztu CEX2020-001058-M,
and Scuola Normale Superiore in Pisa for hospitality during 2020-21.

\bibliographystyle{apalike}
\bibliography{references}

\end{document}